\documentclass[aps,prd,twocolumn,amsmath,showpacs,amssymb,superscriptaddress,nofootinbib,longbibliography]{revtex4-2}

\usepackage{graphicx}
\usepackage{bm}
\usepackage{amssymb,amsmath}
\usepackage{mathrsfs}
\usepackage{latexsym}
\usepackage{mathtools} 
\usepackage{color}
\usepackage[normalem]{ulem} 
\usepackage{dcolumn}

\usepackage[colorlinks=true,citecolor=blue,urlcolor=blue]{hyperref}
\usepackage[usenames,dvipsnames]{xcolor}

\DeclareMathOperator{\sinc}{sinc}
\newcommand{\R}{\tilde{\mathcal R}{}}

\newcommand{\Caltech}{\affiliation{Theoretical Astrophysics 350-17,
    California Institute of Technology, Pasadena, CA 91125}}

\newcommand{\Austin}{\affiliation{Weinberg Institute, University of Texas at Austin, Austin, TX 78712, USA}}

\begin{document}

\title{Rogue echoes from exotic compact objects}

\author{Aaron Zimmerman}
\Austin
\author{Richard N. George}
\Austin
\author{Yanbei Chen}
\Caltech
\date{\today}

\begin{abstract}
Binary systems containing exotic compact objects may emit repeated bursts of gravitational waves (GWs) following coalescence.
Such GW echoes would provide a clear signature of new physics, but searches for them have not yielded a convincing detection.
Here we argue that the typical time delay between a GW event and its echoes is much greater than generally expected, due to long propagation times through objects that mimic black holes.
We provide a simple recipe for computing the time delay and several examples.
These time delays can be billions of years, resulting in rogue echoes that are not correlated with GW events and evade all current constraints.
They would be detectable only by searches for individual echoes or GW bursts.
\end{abstract}

\maketitle

\section{Introduction}
The detection of GWs from compact binaries~\cite{Abbott:2016blz,TheLIGOScientific:2016pea,LIGOScientific:2017vwq,LIGOScientific:2018mvr, Venumadhav:2019tad, Venumadhav:2019lyq, Nitz:2018imz, Nitz:2020oeq, LIGOScientific:2020ibl, Nitz:2021uxj, LIGOScientific:2021usb, LIGOScientific:2021djp,Nitz:2021zwj,Olsen:2022pin} has opened new avenues in the search for physics beyond general relativity and the Standard Model~\cite{Barack:2018yly}.
These discoveries have enabled tests of relativity in strong-field regimes and probed the propagation of GWs, e.g.~\cite{TheLIGOScientific:2016src,Yunes:2016jcc,Carson:2020rea,Abbott:2018lct,LIGOScientific:2020tif,LIGOScientific:2021sio}.
They have also opened up the possibility of discovering entirely new classes of exotic compact objects (ECOs) such as boson stars, dark matter stars, gravastars, and many others~\cite{Cardoso:2019rvt}.
The detection of such objects would reveal physics beyond the Standard Model, modifications to relativity, or even signatures of quantum gravity.
If sufficiently compact, these objects act as black-hole mimickers~\cite{Cardoso:2019rvt,Johnson-McDaniel:2018uvs}, making it a challenge to distinguish these objects from black holes (BHs) using GW observations.

However, because ECOs are not perfect absorbers, they are expected to produce a radically different signal during the ringdown phase following binary merger.
Generally the ringdown modes of these objects have much longer decay times than BHs~\cite{Kokkotas:1994an}.
Studies using perturbation theory around the final state have revealed that generic excitations of this spectrum can lead to a ringdown which initially mimics that of a BH, but which afterward produces a sequence of repeated {\it echoes}, bursts of GWs which gradually give way to ringing at the slowest decaying mode~\cite{Ferrari:2000sr,Cardoso:2016rao,Cardoso:2016oxy}.

The essential idea is that GWs which would be absorbed by the horizon if the merger remnant were a BH are instead reflected back into the spacetime by the ECO, either because the waves propagate into the object and pass through the origin and back out, or because some more exotic mechanism reflects the waves just outside the location of the horizon.
When the remnant is sufficiently compact, these secondary waves are separated in time from the initial merger-ringdown signal, appearing as a sequence of coherent echoes of the merger-ringdown.

The detection of these echoes would provide definitive evidence of the presence of an ECO.
For this reason, there have been extensive efforts to model echoes under various approximations and ECO properties, e.g.~\cite{Cardoso:2016oxy,Mark:2017dnq,Bueno:2017hyj,Nakano:2017fvh,Testa:2018bzd,Maggio:2019zyv,Conklin:2019fcs,Chen:2020htz,Ikeda:2021uvc,Srivastava:2021uku, Wang:2018gin, Micchi:2019yze}, and with methods which can incorporate data from binary merger simulations~\cite{Annulli:2021ccn,Ma:2022xmp,Dailey:2023mvn}.
Searches for these echoes in GW data have been carried out, targeting the data following binary mergers~\cite{Nielsen:2018lkf,Lo:2018sep,Uchikata:2019frs,Tsang:2019zra,Wang:2020ayy,LIGOScientific:2020tif,Ren:2021xbe,LIGOScientific:2021sio}. 
While some have found evidence for echoes~\cite{Abedi:2016hgu,Ashton:2016xff,Abedi:2017isz,Conklin:2017lwb,Abedi:2018pst,Abedi:2020sgg}, others have not,
and there is currently no consensus on the existence of echoes in GW data, see~\cite{Abedi:2020ujo} for a review.
This non-detection of echoes can be used to constrain the reflectivity of BH horizons and the prevalence of ECOs in coalescing binaries~\cite{Cardoso:2017cqb,Cardoso:2017njb,Maggio:2020jml, Maggio:2021ans}. 
In addition, the non-detection of an enhanced stochastic GW background due to echoes~\cite{Du:2018cmp} provides further constraints on the properties of ECOs~\cite{Barausse:2018vdb}.

When searching for echoes after GW events, these studies assume that the typical time delay $\Delta T$ between the echoes is controlled by the propagation time between the photon sphere and the surface of the ECO.
For an object which mimics a BH, its compactness should be nearly that of a BH with the same mass $M$, and so its surface is assumed to lie just outside the horizon radius.
The resulting time delay is $\Delta T \sim 2M \log M/L $, where $L$ is the small distance between the surface of the compact object and the Schwarzschild radius and represents a length scale for new physics~\cite{Cardoso:2016oxy}.
The logarithmic scaling means that $\Delta T$ is of the order of only a few hundred times the light crossing time of the horizon, even for $L$ of order the Planck length.
This expected time scale has been used to target searches for echoes following binary coalescence.

This logarithmic scaling only accounts for the propagation time outside of the object. 
Since the echoes must also pass through the interior of a compact object, additional time delays arise.
For example Pani and Ferrari~\cite{Pani:2018flj} treated the case of an ultracompact Tolman-Oppenheimer-Volkoff (TOV) star when considering the possibility of echoes from the binary neutron star merger GW170817~\cite{TheLIGOScientific:2017qsa}. 
They found that the majority of $\Delta T$ arises from the crossing time through the interior of the star, with a contribution to $\Delta T$ proportional to $\epsilon^{-1/2}$, where
$\epsilon = R/R_B - 1$ and $R_B$ is the Buchdahl radius, $R_B = 9M/4$ (see also~\cite{Urbano:2018nrs}). 
Long time delays have also been noted, for example, for the Damour-Solodukhin wormhole~\cite{Solodukhin:2005qy,Damour:2007ap}, if its deformation parameter is chosen to scale exponentially with the Planck mass~\cite{Damour:2007ap,Sebastiani:2018ktb}. 

These results suggest that the time delays of echoes may be much longer than what has been assumed in past searches.
Instead of following shortly after the merger of compact objects, echoes may appear as rogue bursts of GWs emitted millions or billions of years later, with no correlation to the initial signals.
If this is the case, echoes would have evaded all previous searches as well as constraints from the non-detection of a stochastic GW background~\cite{Du:2018cmp,Barausse:2018vdb}.

In this study, we consider ECOs which are near the threshold of becoming a BH.
In terms of compactness, $\xi = M/R$, we assume the objects have a maximum compactness $\xi_*$ above which the spacetime becomes singular or the interior of the object lies inside of a horizon.
We further assume that this critical compactness is high enough that the photon sphere is exterior to the object, which is necessary for an object to mimic BH ringdown and to produce subsequent echoes.
Defining $\epsilon = \xi - \xi_*$, we give a generic argument that echoes from ECOs with an interior have a delay time that scales as $M \epsilon^{-p}$ with $p \geq 1/2$ when $\epsilon \ll 1$. 
We also provide particular examples, showing that ultracompact TOV stars, thin-shelled gravastars, and Buchdahl stars all have time delays which scale as $\epsilon^{-1/2}$.
We show this using both simple estimates based on the light crossing time of the object, as well as a more refined WKB analysis of the time delay.
Our WKB analysis is similar to that of~\cite{Cardoso:2014sna,Sebastiani:2018ktb}.
We also discuss the detection prospects for rogue echoes.
In this study we use geometric units, with $G = c = 1$.

\section{Time delays}

As a model for our ECO, consider the static, spherically symmetric spacetime of a perfect fluid star with radius $R$. 
We use the areal radius $r$ and a Killing time $t$ as our coordinates, so that
\begin{align}
    ds^2 & = - g_{tt} dt^2 + g_{rr} dr^2 +  r^2 d\Omega^2 \,, \\
    g_{rr} & = \left(1- \frac{2m(r)}{r}\right)^{-1} \,, 
    \qquad
    m(r)  \coloneqq 4 \pi \int_0^r \rho(r)r^2 dr \,,
\end{align}
where $\rho$ is the energy density of the matter. 
The light-crossing time across the interior of the ECO is
\begin{align}
\Delta T = 2 \int_0^R \frac{dt}{dr} dr = 2 \int_0^R \sqrt{\frac{g_{rr}}{-g_{tt}} } dr \,.
\end{align}

Generically, ECO spacetimes which exceed a certain compactness $\xi_*$ become singular or else form a horizon somewhere in the spacetime, which in these coordinates occurs when $g_{tt} = 0$ at some radius.
For simplicity, consider the case where the ECO remains regular when $R > 2M$, so that $\xi_*= 1$, and at the critical compactness a horizon forms at the Schwarzschild radius.
In order to match the interior to the exterior spacetime at $r=R$, we have $|g_{tt}(R)| = 1- 2M/R = \epsilon$. 

The essential idea is that this small value of the lapse function $|g_{tt}|$ near the surface results in a small value the lapse through an extended region inside the star, and so a slow propagation through the star.
The TOV equations~\cite{Schutz:1985jx} show that for positive $\rho$ and pressure $P$, $|g_{tt}|$ decreases towards the center of the star, so that $0 < |g_{tt}| \leq \epsilon$ inside the star.
Meanwhile, $g_{rr}(R) = \epsilon^{-1}$ but falls rapidly towards $1$ as we move to lower radii, since $m(r)$ tends to scale as $r^k$, $k\geq 3$.
We can conservatively bound $1 \leq g_{rr} \leq \epsilon^{-1}$ inside the star. 
Taken together, we find
\begin{align}
    \Delta T \geq 2 R \epsilon^{-1/2} \,.
\end{align}
This time delay $\sim M \epsilon^{-1/2}$ is much longer than the contribution from the propagation of the signal from the light ring to the surface of the star, $\sim M \ln \epsilon$.

Since we are considering exotic objects it is reasonable that one or both of $\rho$ or $P$ become negative within the interior of the star, breaking our assumptions. 
Even in this case, there will be a region $\Delta R$ where $|g_{tt}|\sim \epsilon^{-1}$ and $g_{rr}$ is large and decreasing towards the interior, which provides a contribution $\Delta T \sim \Delta R \epsilon^{-q}$ with $1/2 \lesssim q \lesssim 1$.

The expected size of $\epsilon$ is uncertain, and depends on the ECO model.
If the formation of a BH is prevented by quantum effects we may expect that new physics arises at the Planck scale, so that $\epsilon \sim \ell_{\rm Pl}/M$, with $\ell_{\rm Pl}$ the Planck length.
Assuming instead that new physics enters at a length scale $\ell \gtrsim \ell_{\rm Pl}$, so $\epsilon \sim \ell/M$, the time delay between echoes is
\begin{align}
\label{eq:ScalingWithNumbers}
    \Delta T \gtrsim 3.1  \left(\frac{\ell_{\rm Pl}}{\ell} \right)^{1/2} \left(\frac{M}{65 M_\odot} \right)^{3/2}{\rm Gyr}\,.
\end{align}
Thus, echoes from stellar mass binary BHs would be observed long after the initial GW signal, but before a Hubble time has elapsed. 
Meanwhile, intermediate mass BH binaries have $\Delta T$ larger than a Hubble time, unless the energy scale of new physics is smaller than $m_{\rm Pl}$.

Our argument that $\Delta T \sim M \epsilon^{-1/2}$ is generic, but involves several assumptions. 
We now discuss three ECO models, all of which display this time delay scaling even when some of our assumptions are violated.
Details are given in 
Appendix~\ref{sec:Supp}.
Note that models which omit an interior entirely, for example involving direct reflection from a surface outside the horizon or wormhole models that match two surfaces outside the horizon~\cite{Damour:2007ap,Cardoso:2016oxy,Cardoso:2016rao,Bueno:2017hyj}, retain the scaling $\Delta T \sim M \ln \epsilon$.

\subsection{Buchdahl's star}

\begin{figure}[tb]
\includegraphics[width=0.98\columnwidth]{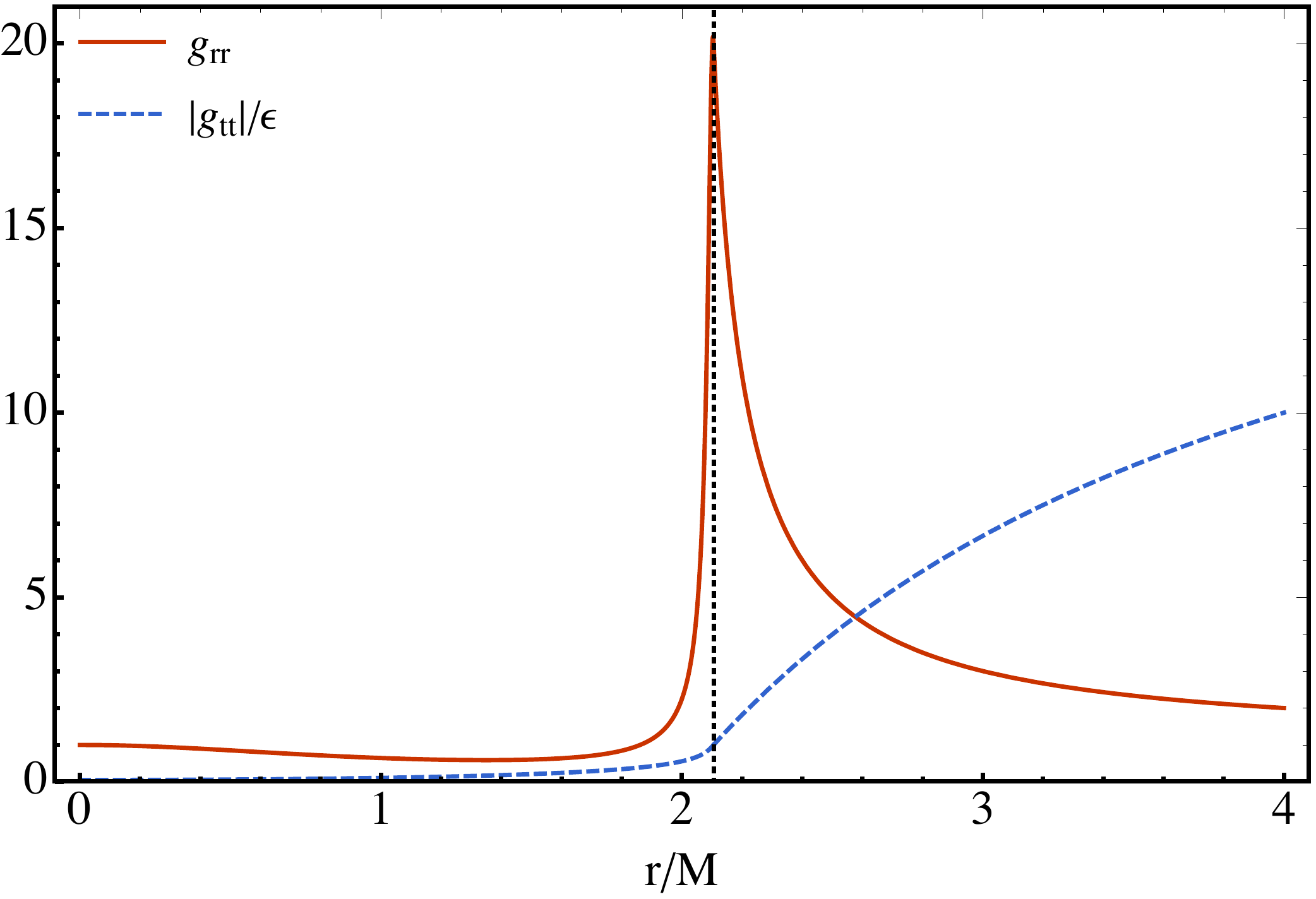} 
\caption{
The metric functions for Buchdahl's star, taking $\epsilon = 0.05$. 
The lapse $|g_{tt}|$ is rescaled by $\epsilon^{-1}$ to increase its visibility, and remains small and decreasing through the star. 
The radial metric function drops sharply from its peak value $g_{rr} \sim \epsilon^{-1}$ inside of the stellar surface $R$, and the time delay $\Delta T$ is dominated by the behavior of the lapse.}
\label{fig:BuchdahlFuncs}
\end{figure}

As a representative example of an ECO spacetime, consider Buchdahl's exact interior solution, described in e.g.~\cite{Schutz:1985jx}.
The metric functions $g_{tt}$ and $g_{rr}$ are depicted in Fig.~\ref{fig:BuchdahlFuncs} with convenient rescalings, using a somewhat large value $\epsilon = 0.05$.
This solution is found with an equation of state $\rho = K^{-1/2} P^{1/2} - 5 P$, which limits to an $n = 1$ polytrope for small pressure $P$.
The solution is parametrized by the constant $K$ (which determines the mass $M$) and the compactness $\xi$ which has a maximum of $\xi_* = 1$.
When $\xi = 1 - \epsilon$ for $\epsilon \ll 1$, the interior star has unphysical characteristics such as a superluminal sound speed and a negative energy density throughout a region around the center of the star.
For this extreme case, we find that $|g_{tt}|\sim \epsilon^2$ in a small region of size $\sim (\epsilon R)^{1/2}$ around $r = 0$.
This region dominates the integral for $\Delta T$, as can be seen in the scaling solution leading to $\Delta T \approx 8M (3/\epsilon)^{1/2}$.
This approximation provides good agreement with numerical results.

\subsection{Gravastar}

A thin-shell gravastar~\cite{Mazur:2001fv,Visser:2003ge,Pani:2009ss} provides a simple model for an ECO which illustrates our generic argument for the scaling of $\Delta T$.
A gravastar has a thin shell of matter surrounding an interior de Sitter solution. 
The total mass $M$ has contributions from both the surface, $M_s$ and the interior volume, $M_v$.
The lapse $-g_{tt}$ is matched at the surface to the exterior, and then grows to $-g_{tt} = (R - 2M)/(R-2M_v)$ at the center of the star.
The maximum compactness is given by $\xi_* = 1$, and the integral for $\Delta T$ can be given in closed form. 
For $\epsilon = 1- \xi$, we find at leading order
\begin{align}
\label{eq:GravastarDelay}
    \Delta T \approx 
    \frac{2M}{\sqrt{\epsilon}} 
    \sqrt{\frac{M-M_v}{M_v}} \tanh^{-1}\sqrt{\frac{M_v}{M}}\,,
\end{align}
provided $M_v/M \gg \epsilon$, so that the surface density represents a non-negligible amount of mass.

\subsection{Constant density TOV star}

The case of a constant density TOV star~\cite{PhysRev.55.364,PhysRev.55.374,Schutz:1985jx} provides an especially simple model for an ECO.
In this case, the critical compactness where collapse to a BH is required occurs when the star has the Buchdahl radius, $R = 9M/4$, so that $\xi_* = 8/9$ rather than $1$.
At this compactness, $g_{tt}(r=0) \to 0$.
Although the details are different than our generic argument above, and $|g_{tt}|$ is not constrained to be small by matching, we nevertheless find the same scaling behavior as for Buchdahls' star.
The result is that $|g_{tt}|\sim \epsilon^2$ in a region of size $\sim (\epsilon R)^{1/2}$ around $r = 0$. 
Again we can compute the integral at leading order and find $\Delta T \approx 10M \epsilon^{-1/2}$.

\section{Effective reflectivity of compact objects}

So far we have discussed the light-crossing time of the interior of ECOs.
These results carry directly through to the propagation of waves through ECOs, and allow us to compute the effective reflectivity $\R(\omega)$~\cite{Mark:2017dnq} of ECOs near the BH threshold.
This reflectivity allows for the construction of simple echo waveforms, where the waves impinging on the ECO are reflected from the ECO with a change in amplitude and a time delay.
For simplicity we focus on the scalar wave equation, which captures the qualitative features of a GW analysis, especially in the WKB limit, e.g.~\cite{Yang:2012he}.

Within an ECO, the scalar wave equation $\Box \psi = 0$ separates into frequency and spherical harmonics, reducing to an equation governing the radial wavefunction $u_{\ell m \omega}(r)$,
\begin{align}
    \frac{d^2 u_{\ell m \omega}}{dr_*^2} + (\omega^2 - V) u_{\ell m \omega} = 0 \,.
\end{align}
The potential 
\begin{align}
V = (-g_{tt})\frac{\ell(\ell+1)}{r^2} + \frac{1}{2r} \frac{d(-g_{tt}/g_{rr})}{dr}   
\end{align} 
tends to be small and vary relatively slowly through the star for ECOs on the threshold of collapse until a centrifugal barrier at small $r$ is reached. 
The tortoise coordinate $r_*$ is the integral of the time delay factor for null rays, $dr_*/dr = (-g_{rr}/g_{tt})^{1/2}$, and is stretched out compared to the areal radius $r$.

The effective potential reflects waves with low frequencies, $\omega \lesssim \omega_{\rm QNM}$, with $\omega_{\rm QNM}$ the fundamental quasinormal mode frequency. 
As such these waves do not reach the ECO or contribute to the echoes.
We instead focus on higher frequencies. 
Because $V(r)$ has a peak near the photon sphere and decreases steeply towards the ECO surface, we expect $\omega^2 \gg V$ near the surface, and by continuity into the ECO interior. 
In fact for the examples we study in this work $\omega \gg V(r)$ when $\omega \gtrsim \omega_{\rm QNM}$, and $V(r)$ varies slowly through much of the ECO interior.
This justifies the use of a standard WKB analysis, which gives an approximate solution for $u_{\ell m \omega}$ in the ECO.
Matching this to the exterior solution allows us to compute the effective reflectivity $\R$, including the phase delay due to propagation through the ECO interior. 

\begin{figure}[t!]
    \centering
    \includegraphics[width = .98 \columnwidth]{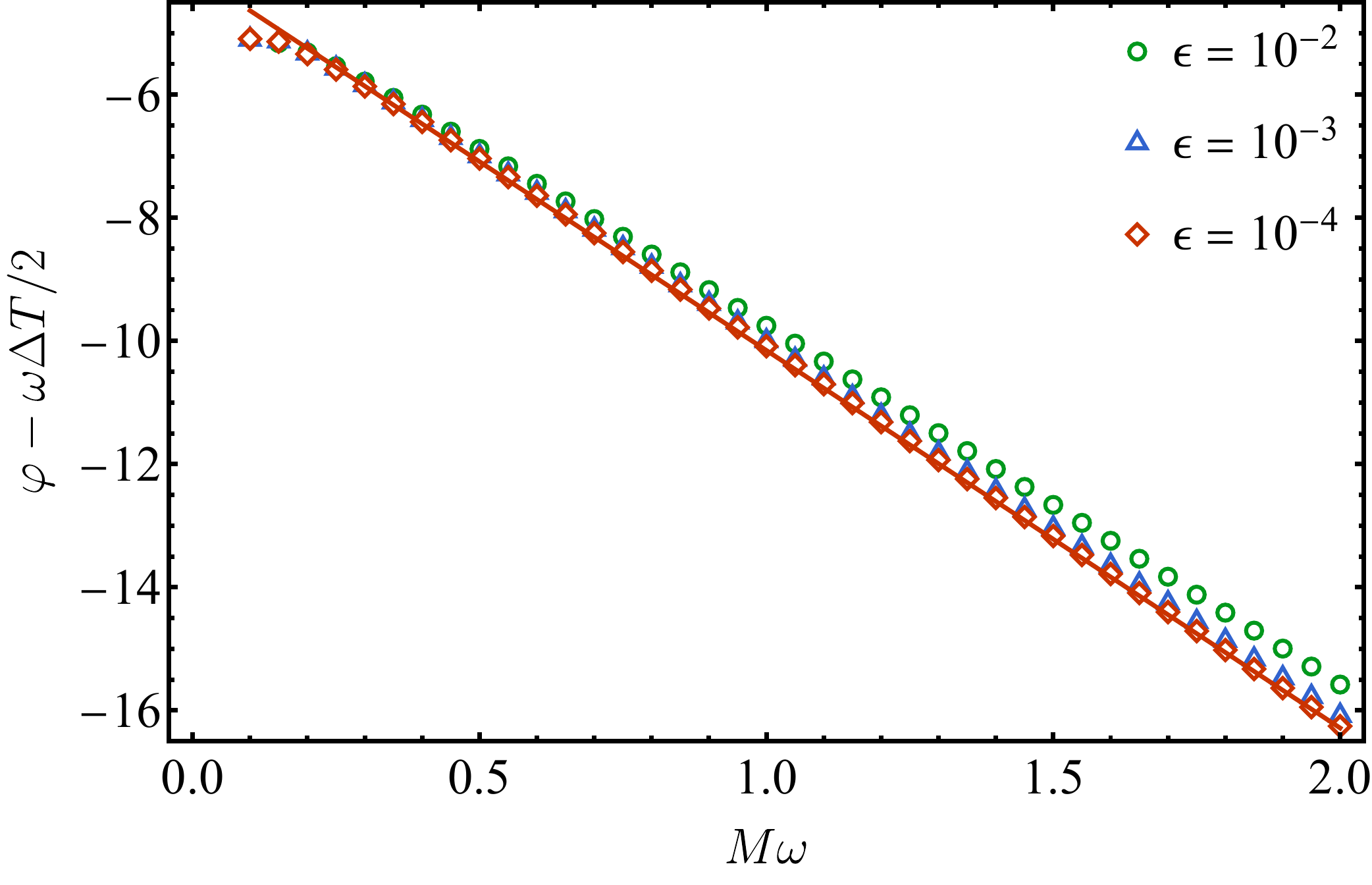}\\
    \caption{ Three iterations of the phase with decreasing values for $\epsilon$: $10^{-2}$, $10^{-3}$, and $10^{-4}$. These points are modeled well by the linear Eq.~\eqref{eq:linearphi}, which indicates there is no dispersion for frequencies $M \omega \gtrsim 0.15$.}
    \label{fig:Dispersion}
\end{figure}

Under the simplifying assumption that the ECO has a maximum compactness of $\xi = 1$, we find
\begin{align}
\label{eq:ECORef}
    \R & = e^{2 i \varphi_R}e^{-i \pi/2}\,, \\
    \varphi_R & = \omega \int_{r_0}^R \sqrt{\frac{g_{rr}}{-g_{tt}}\left(1 - \frac{V}{\omega^2}\right)} \, dr \,.
\end{align}
Here $r_0$ is the classical turning point of the wave due to the centrifugal barrier near the origin, $\omega^2 = V(r_0)$.
This result shows that the reflection is total, and induces a phase delay equal to twice $\varphi_R$ (the phase evolution from the surface to the turning point) plus the factor $-\pi/2$ arising from the WKB matching conditions.
In the approximation that $r_0 \approx 0$ and $V$ is negligible in the interior, we recover $2 \varphi_R \approx \omega \Delta T$, validating our simple estimates.
The contribution of $r_0$ and $V$ make $\R$ frequency dependent, and can lead to mild dispersion of waves traveling through the ECO.
We illustrate this by comparing $\varphi_R/\omega $ to the leading approximation for $\Delta T$ for Buchdahl's solution in Fig.~\ref{fig:Dispersion} in the case of small $\epsilon$. We find
\begin{align}
\label{eq:linearphi}
     \varphi_R \approx \frac{4\sqrt{3}}{\sqrt{\epsilon}}\omega + (a+b\omega) ,
\end{align}
where $a = -4.02$ and $b = -6.13$.
At this level of approximation, there is no dispersion of the waves which make up the echo.
This means that methods based on the simple recipe for echoes in Ref.~\cite{Mark:2017dnq}
can be applied with a simple adjustment to the delay time between the echoes.
Greater detail on the WKB approximation used here is given in Appendix~\ref{sec:SuppWKB}.

\section{Discussion}

So far we have argued that long delay times $\Delta T \gtrsim {\rm Gyr}$ between echoes are a generic consequence of the deep potential well and strong time dilation within ECOs.
This has a number of consequences.
First, such long delay times mean that 
there would be no echoes visible following a GW signal within the current observational timescale.
Thus targeted searches for echoes following GW events would not detect echoes from ECOs with interiors, even if such objects were mimicking BHs.
Rather than searching for echoes correlated with detected GWs, we would need to search for rogue echoes, GW bursts with similar duration and frequency content to merger-ringdown signals from binary coalescence, but whose morphology is altered by reflection and transmission in the cavity inside the effective potential~\cite{Mark:2017dnq}.

Long delay times also weaken the constraints on ECOs from the non-detection of a stochastic GW background.
In Ref.~\cite{Barausse:2018vdb}, the stochastic background due to echoes and instabilities in rotating ECOs is computed, both in the case where $\Delta T \sim M \ln \epsilon$ and in the case where the delay time is arbitrary due to uncertain propagation time through the interior.
Since GWs sum incoherently when producing the stochastic background, the strain of the background scales as $1/\sqrt{N}$ where $N$ is the number of emitting sources, and the energy density scales as $1/N$. 
By increasing the time delay between echoes, the number of sources (in this case, echoing bursts) decreases proportionately, reducing the background.
For our fiducial parameters in Eq.~\eqref{eq:ScalingWithNumbers}, $\Delta T \sim 10^{20} M$, many orders of magnitude above the value $\Delta T \sim 10^8 M$ which can be constrained by Advanced LIGO and Virgo~\cite{Barausse:2018vdb}.

If echoes appear as single bursts, the need for accurate models of echoes is more pressing.
Matched filtering with theoretical templates would provide the most sensitive searches for single echoes, but modeling echoes presents a number of serious theoretical and technical challenges.
For one, ECOs which spin are susceptible to ergoregion instabilities, see e.g.~\cite{Friedman:1978,Cardoso:2007az,Cardoso:2008kj,Cardoso:2014sna,Brito:2015oca,Maggio:2017ivp,Maggio:2018ivz,Zhong:2022jke}, which pose difficulties for creating realistic background models. 
Secondly, perturbed ECOs may be susceptible to the formation of an enclosing horizon~\cite{Chen:2019hfg}.

Formulating a model for the dynamical formation of ECOs compact enough to produce echoes following binary merger is even more daunting. 
Numerical analyses of currently tractable models such as boson stars and soliton stars~\cite{Helfer:2018vtq} reveal the challenges in producing a sufficiently compact final state.
However the results of~\cite{Helfer:2018vtq}, whose sub-critical head on collisions display echoes with $\Delta T \sim 600 M$, are promising.
Simulations like these are needed to produce truly realistic signal models for ECOs.

Long propagation times in the interior of ECOs impact a number of other predicted GW signatures.
For example, resonant excitation of the modes of ECOs during during extreme-mass-ratio inspirals could provide observable effects on the signal~\cite{Cardoso:2019nis,Maggio:2021uge,Sago:2021iku}.
If the characteristic timescale of the modes, and thus the required timescale to build up the resonant mode amplitude, is very long then resonant effects would be negligible (see e.g.~\cite{Cardoso:2022fbq}).

Currently, the strongest constraints on rogue echoes with long time delays are likely to come from minimally modeled, all-sky searches for GW bursts~\cite{Drago:2020kic,Cornish:2020dwh}.
For the third observing run of the LIGO and Virgo detectors, these searches reported a root-mean-square amplitude sensitivity of $h_{\rm rss} \sim 10^{-22}$ Hz${}^{-1/2}$~\cite{KAGRA:2021tnv}.
This roughly corresponds to the $h_{\rm rss}$ of a $65 M_\odot$ binary BH at 1 Gpc when integrated across the sensitive frequency band of two ground-based detectors.
Although only a fraction of the signal would contribute to the an echo and the amplitude of the echo is likely further suppressed, this nevertheless indicates that current unmodeled searches may constrain ECOs.
For such constraints, the rate density of binary coalescence must be convolved with our delay time estimates and with simple echo templates.
We leave such investigations, as well as tailored searches for rogue echoes, to future work.

\acknowledgements

We thank Vitor Cardoso, Elisa Maggio, and Paolo Pani for valuable discussions.
AZ is supported by NSF Grant PHY-2308833.
RNG is supported by a Provost Graduate Fellowship at UT Austin.
YC is supported by the Brinson Foundation, the Simons Foundation (Award Number 568762), and NSF Grants PHY-2011961 and 2011968.
This work has been assigned preprint number UTWI-22-2023.

\appendix

\section{Detailed time delay calculations}
\label{sec:Supp}

\subsection{TOV equations}

Let $g_{tt} = - e^{2 \Phi(r)}$ and $P(r)$ be the pressure, then the TOV equations are
\begin{align}
    \frac{d\Phi}{dr} & = \frac{m(r) + 4 \pi r^3 P(r)}{r(r-2m(r))}\,, \\
    \frac{dP}{dr} & = - (\rho + P)  \frac{d\Phi}{dr} \,.
\end{align}
These equations must be supplemented with an equation of state relating pressure $P$ to the energy density $\rho$.
The surface of the star is defined through $P(R) = 0$ and $g_{tt}$ is matched to the exterior there.
We see that $d\Phi/dr|_R > 0$ for $\xi >1$ and $M>0$.
As we move from $R$ inward, $|g_{tt}|$ decreases further, and pressure increases.
This causes the rate of propagation $dt/dr$ to decrease as we move inward.
These trends can be reverse if $\rho$ becomes negative, subject to the global condition $M>0$, at some radii inside the star.

\subsection{Buchdahl's star} 

This solution is described in \cite{Schutz:1985jx}, although we adopt some modifications to the notation there.
The metric functions are
\begin{align}
g_{tt} & = - (1-\xi)\frac{2 - \xi - \xi \sinc z}{2 - \xi + \xi \sinc z } \,,\\
g_{rr} & =\frac{4 (1- \xi)}{(2 - \xi + \xi \cos z)^2}\frac{2 - \xi + \xi \sinc z}{2 - \xi - \xi \sinc z}\, 
\end{align}
The coordinate $z$ is related to the areal radius $r$ via
\begin{align}
\label{eq:rofz}
    r(z) = \frac{2Mz(2 - \xi + \xi \sinc z)}{\pi \xi (2- \xi)}\,.
\end{align}
The boundary of the star is at $z = \pi$. 
For Buchdahl's star we see the maximum compactness occurs at $\xi = 1$. 
The lapse $|g_{tt}|$ vanishes when $\xi \to 1$, as expected from the matching with the exterior spacetime at this extreme compactness.

\begin{figure}[t]
\includegraphics[width=0.98\columnwidth]{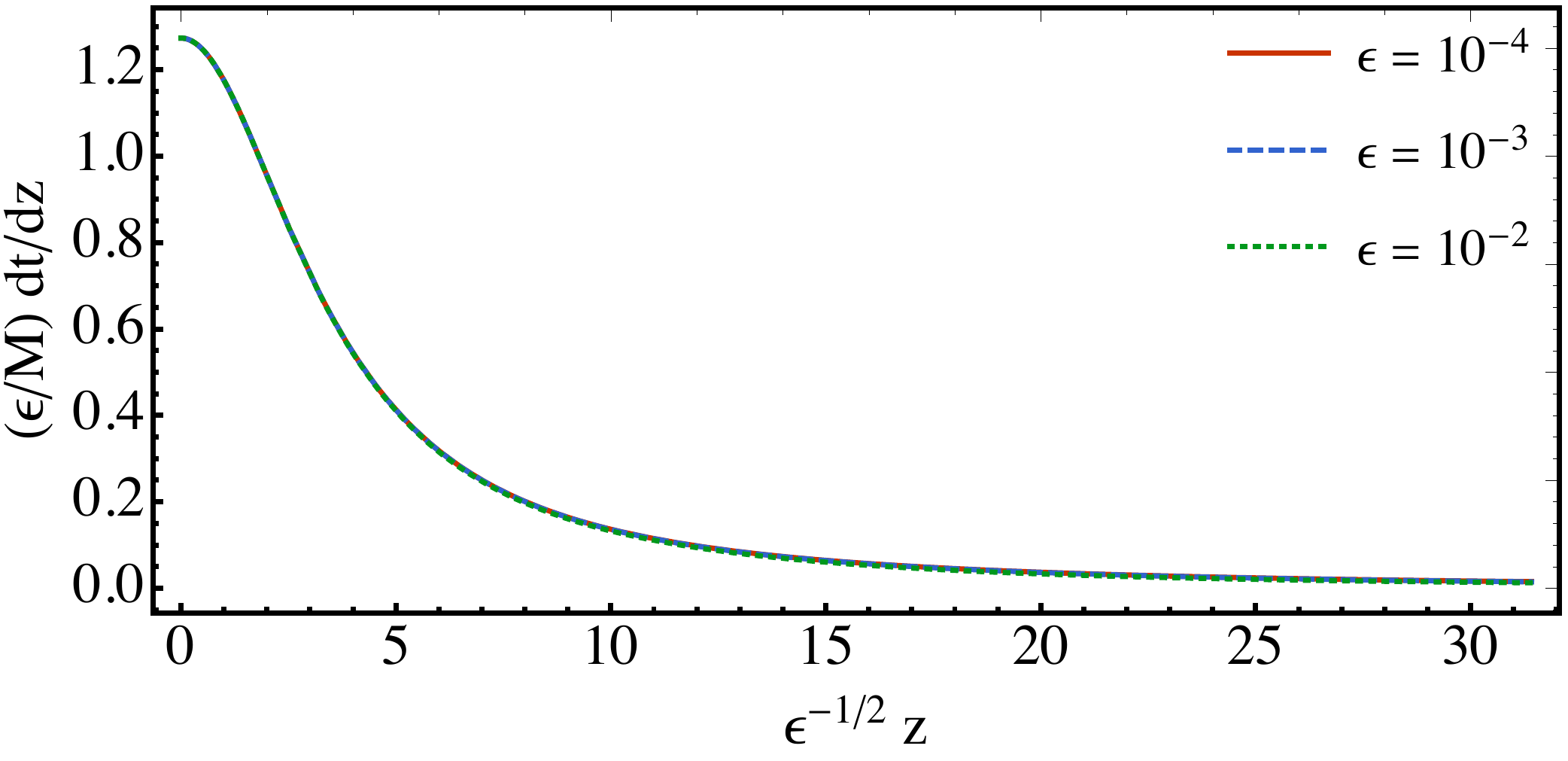}\\
\includegraphics[width=0.98\columnwidth]{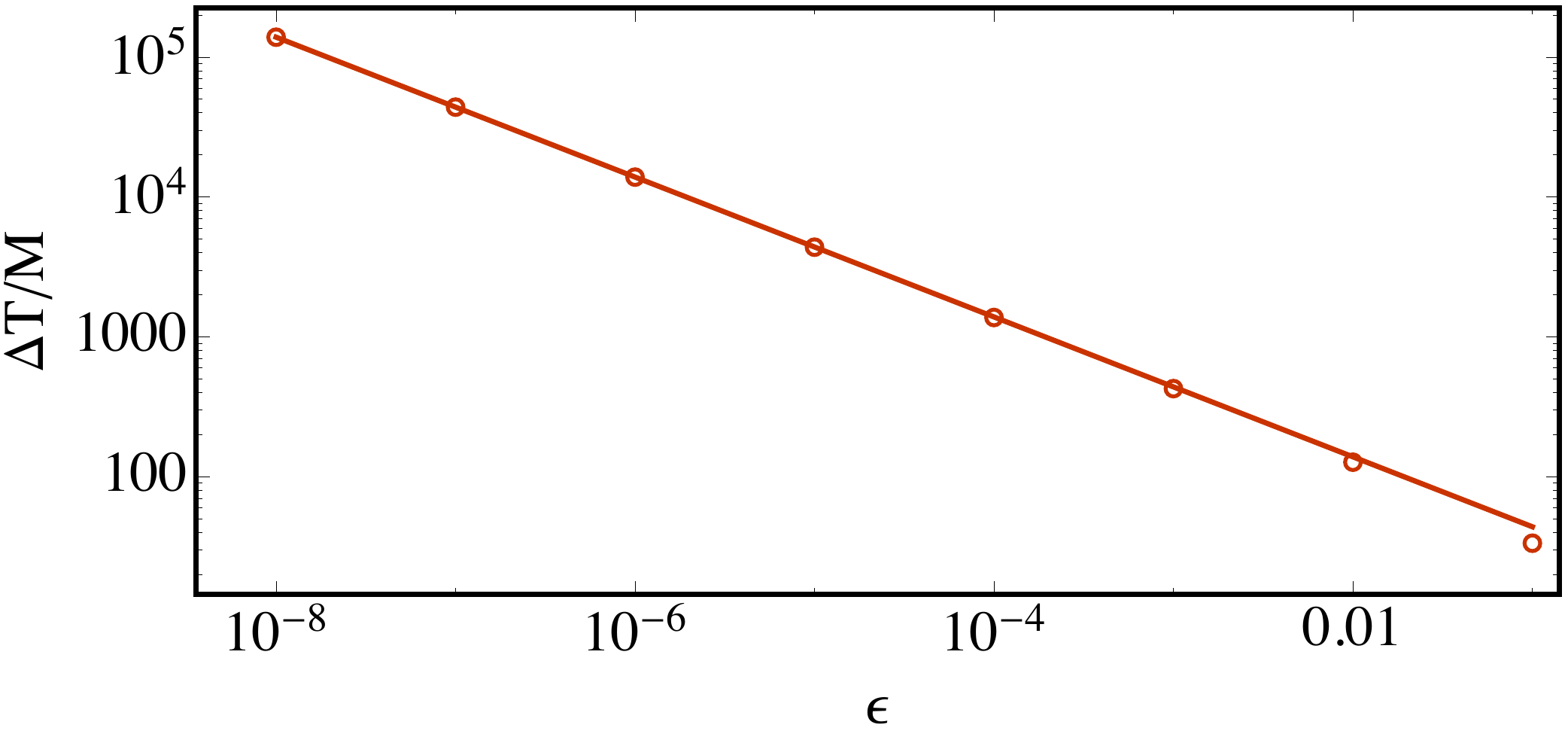}
\caption{{\it Top panel}: Rescaled time delay increment $(\epsilon/M) dt/dz$ for Buchdahl's interior solution, plotted in terms of a rescaling of the radial coordinate for several values of $\epsilon$. The greatest contribution to $\Delta T$ comes from the narrow region near the origin, and motivates our approximate solution Eq.~\eqref{eq:BuchdahlInt}.
{\it Bottom panel}: Numerical evaluations for $\Delta T$ for Buchdahl's solution, as a function of $\epsilon$. The numerical values (open circles) are in good agreement with the analytic approximation (solid line) from Eq.~\eqref{eq:BuchdahlInt}.
}
\label{fig:RescaledBuch}
\end{figure}

The time delay factor for $\epsilon = 1-\xi \ll 1$ is
\begin{align}
    \Delta T & = 2 \int_0^\pi \sqrt{\frac{g_{rr}}{-g_{tt}}} \frac{dr}{dz} dz 
    \notag \\ 
    & = \frac{8M}{\pi} 
    \int_0^\pi \frac{1 + \sinc z}{1 + \epsilon - (1-\epsilon) \sinc z}  dz\,.
\end{align}
When $z \to 0$ the integrand diverges as $\epsilon^{-1}$, but outside of a narrow range of small $z$ values the integral is $O(1)$.
The region over which the integrand is $\sim \epsilon^{-1}$ is $\sim \epsilon^{1/2}$, which can be observed in the top panel of Fig.~\ref{fig:RescaledBuch}.
The universal dependence of $\epsilon \, dt/dz$ with respect to $\epsilon^{-1/2} z$ as $\epsilon$ varies motivates the definition of a coordinate $y = \epsilon^{-1/2} z$ and splitting the radial integral into two parts: 
one over a region $0 \leq y \leq Y$ with $Y \ll \epsilon^{1/2}$ so that $\epsilon \, y^2$ can be considered small throughout the domain of the integral, and another covering the region from $ \epsilon^{1/2} Y <z\leq\pi$.
At leading order, the integral over the first region dominates, and we find
\begin{align}
\label{eq:BuchdahlInt}
    \Delta T 
    & \approx \frac{8 M}{\pi\sqrt{\epsilon}} \int_0^Y \frac{dy}{1+y^2/12} 
    \approx \frac{8 M}{\pi\sqrt{\epsilon}} \int_0^\infty \frac{dy}{1+y^2/12}
    \notag \\ 
    & \approx  8M \sqrt{\frac{3}{\epsilon}} \,,
\end{align}
where we have extended the integration limit to infinity to get a closed form expression, relying on the fact that the integrand is small beyond $Y$.
This final result is in excellent agreement with a fit to numerical integrals for $\Delta T$ using a sequence of small $\epsilon$ values; these numerical values are compared to the analytic approximation in the bottom panel of Fig.~\ref{fig:RescaledBuch}.

\subsection{Gravastar}

For the gravastar model~\cite{Pani:2009ss}, it is useful to define $\beta = 2M_v/R$ in addition to the compactness $\xi$ and a scaled radius $x = r/R$. In terms of these, metric functions are
\begin{align}
    g_{tt} & = -\frac{1-\xi}{1 - \beta} (1 - \beta x^2)\,. \\
    g_{rr} & = (1- \beta x^2)^{-1}\,.
\end{align}
From these, we find
\begin{align}
    \Delta T = 2R\sqrt{\frac{1-\beta}{\epsilon}} \int_0^1 \frac{dx}{1- \beta x^2} = 2R\sqrt{\frac{1-\beta}{\beta \epsilon}}\tanh^{-1} \sqrt \beta \,.
\end{align}
When $\epsilon \ll1$ and $1-\beta \gg \epsilon$ we recover Eq.~\eqref{eq:GravastarDelay}. If this is not the case then some care is needed. For example if $M_s = 0$, then $\Delta T$ has a log divergence rather than $\sim \epsilon^{-1/2}$.

\subsection{TOV star}
\label{sec:TOVdetails}

\begin{figure}[t!]
    \centering
    \includegraphics[width = 0.98\columnwidth]{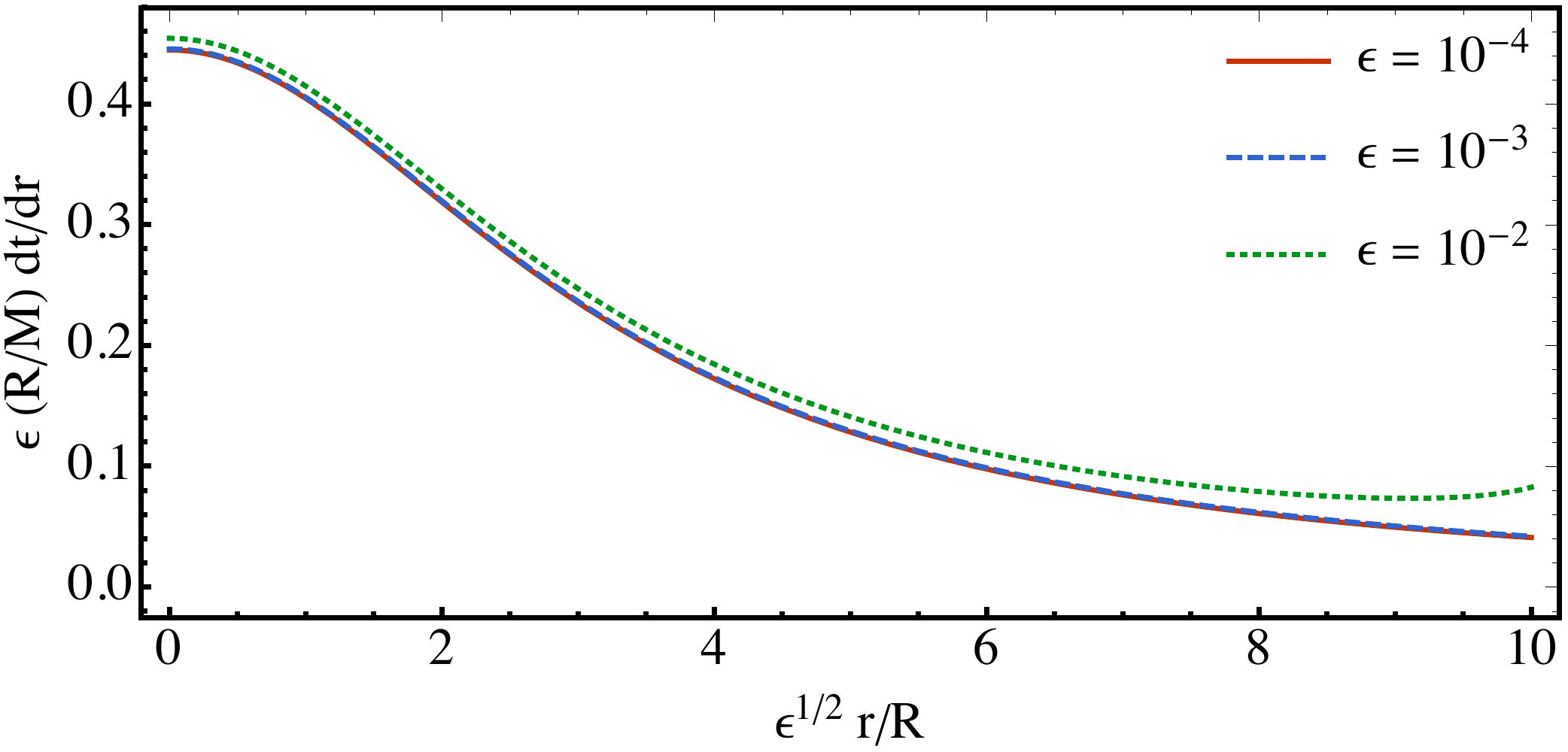}\\
    \includegraphics[width = 0.98\columnwidth]{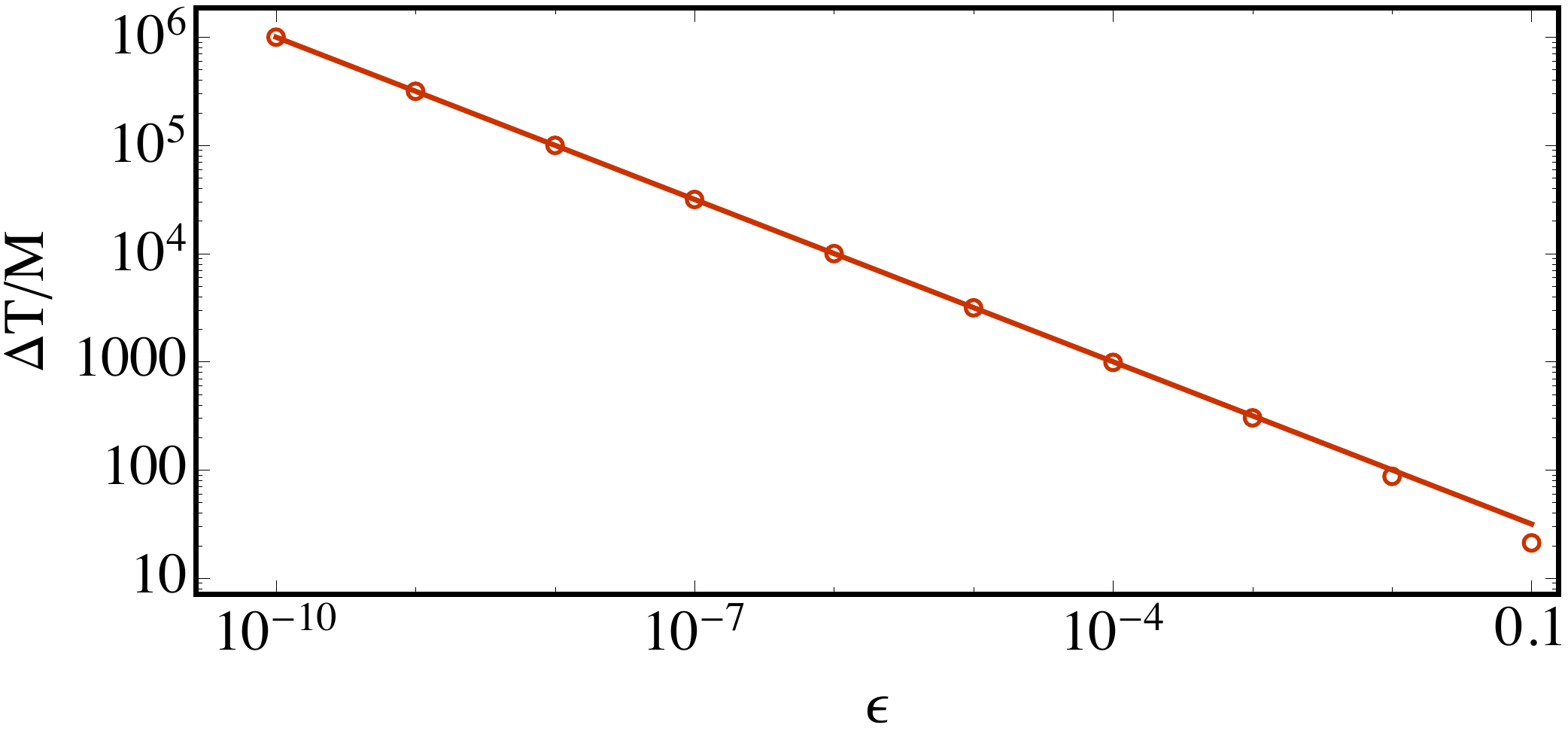}
    \caption{{\it Top}: Scaled inverse radial velocity within the TOV star, showing the universal scaling of the light crossing time at small $\epsilon$.
    {\it Bottom}: Numerical evaluations for $\Delta T$ for Buchdahl's solution, as a function of $\epsilon$. The numerical values (open circles) are in good agreement with the analytic approximation (solid line) from Eq.~\eqref{eq:TOVInt}}
    \label{fig:RescaledTOV}
\end{figure}

When $\rho$ is constant, we immediately have $M = 4\pi \rho R^3/3$ and $m(r) = M x^3$, where we define $x = r/R$. We have
\begin{align}
    g_{tt} & = - \frac{1}{4} \left(3 \sqrt{1 - \xi} - \sqrt{1 - \xi x^2} \right)^2 \,,
\end{align}
which vanishes at $x = 0$ when $\xi = \xi_* = 8/9$.
The central pressure, $P_c = P(x = 0)$, blows up at this compactness, with $P_c \propto \epsilon^{-1}$ for $\epsilon = \xi - \xi_*$.
For stars with $\epsilon \ll 1$, $|g_{tt}|$ is only nearly zero in a region around $r = 0$, where $|g_{tt}|\sim \epsilon^2$ is fairly flat before increasing to match the outer solution.
The region over which $|g_{tt}| \sim \epsilon$ is $\sim \epsilon^{1/2}$ in size, which can be observed in the top panel of Fig.~\ref{fig:RescaledTOV}.
Meanwhile, $g_{rr} = (1- \xi x^2)^{-1}$ ranges in value between 1 and 9 for the most extreme case, and does not play a role in the scaling of $\Delta T$ with $\epsilon$.

The universal behavior of $g_{tt}$ motivates the definition of a coordinate $y = x \epsilon^{-1/2}$ just as for the Buchdahl star, with the same splitting of the integral: one over a region $0 \leq y \leq Y$ with $Y \ll \epsilon^{1/2}$ so that $\epsilon y^2$ can be considered small throughout the domain of the integral, and another covering the region from $ \epsilon^{1/2} Y <x\leq1$.
At leading order we get
\begin{align}
\label{eq:TOVInt}
    \Delta T & \approx\frac{8R}{9\sqrt{\epsilon}}\int_0^Y \frac{dy}{1+8y^2/81} 
    \approx \frac{8R}{9\sqrt{\epsilon}}\int_0^\infty \frac{dy}{1+8y^2/81} 
    \notag \\
    & = \frac{4\pi M}{\sqrt{2\epsilon}}\,,
\end{align}
using the same logic as in Eq.~\eqref{eq:BuchdahlInt}.
The rescaled integrand $\epsilon^{1/2} dt/dx$ is plotted as a function of $y$ in the bottom panel of Fig.~\ref{fig:RescaledTOV}, illustrating the accuracy of the leading order approximation.
This final result is in excellent agreement with a fit to numerical integrations.

\pagebreak
\onecolumngrid

\section{Detailed WKB analysis}
\label{sec:SuppWKB}

Here we present further details on our WKB computations for ECOs.

\subsection{Scalar wave equation inside ECO}

We begin with the homogeneous scalar wave equation
\begin{align}
    \Box_g \Psi = \frac{1}{\sqrt{-g}}\partial_\mu (\sqrt{-g} g^{\mu \nu} \partial_\nu \Psi )= 0
\end{align}
and apply separation of variables with $\Psi = e^{-i\omega t} R_{\ell m \omega}(r) Y_{\ell m}(\theta,\phi)$.
The result is a radial wave equation
\begin{align}
    r^{-2} \sqrt{\frac{f}{h}}\frac{d}{dr}\left( r^2 \sqrt{\frac{f}{h}} \frac{dR_{\ell m \omega}}{dr}\right) + \left(\omega^2 - \frac{f\ell(\ell+1)}{r^2}\right)R_{\ell m \omega} = 0
\end{align}
with $g_{tt} = - f$, $g_{rr} = h$\,.
With a change of variables $R_{\ell m \omega} = u_{\ell m \omega}(r)/r$ we rewrite this as
\begin{align}
    \frac{d^2 u_{\ell m \omega}}{dr_*^2} + \left[\omega^2 - V(r)\right] u_{\ell m \omega} = 0 \,.
\end{align}
Here,
\begin{align}
    \frac{dr_*}{dr} & = \sqrt{\frac{h}{f}} = \left. \frac{dt}{dr}\right|_{\rm null\, ray}\,, & 
    V(r) & = f\frac{\ell(\ell+1)}{r^2} + \frac{1}{2r}\frac{d}{dr}\frac{f}{h} \,.
\end{align}
Note that in the exterior, we have $h = 1/f$ and $f = 1- 2M/r$, so that $r_*$ agrees with the usual tortoise coordinate and we can also see that the wave equation agrees with the standard one in Schwarzschild for $u$, e.g.~\cite{Berti:2009kk}.

Inside the ECO we are in a regime where $V(r)$ varies slowly until we reach a steep centrifugal barrier near the origin at $r = 0$. 
The problem is one of a wavefunction with energy $\omega^2$ propagating in one dimension: when $\omega^2 < V(r)$ the wave becomes evanescent and decays exponentially.
Meanwhile, at the surface of the star the problem switches over to one of free propagation, since the spacetime outside of a very compact star is that of the near-horizon region of a BH.

Following \cite{Dalsgaard:2003book}, we write the standard WKB solution inside the star as
\begin{align}
\label{eq:WKBWaveform}
    u_{\ell m \omega} & = \frac{A}{(\omega^2 - V)^{1/4}} \cos [\varphi(r_*) - \pi/4] \,, 
    & \varphi(r_*)= \int_{r_{*,0}}^{r_*} \sqrt{\omega^2 - V(r)} dr_* \,,
\end{align}
where $r_{*,0}$ is the radius where $\omega^2 = V$ (the turning point for classical motion in this potential).
This solution is derived in the WKB limit by matching to the decaying solution beyond the centrifugal barrier near $r = 0$.
The WKB approximation is valid in the region where $(d\varphi/dr_*)^2 \gg |d^2 \varphi/dr_*^2|$, and in this case we see
\begin{align}
    \beta &\coloneqq \frac{|d^2 \varphi/dr_*^2|}{(d\varphi/dr_*)^2} = \frac{|(dV/dr)(dr/dr_*)|}{\omega^2 - V} \gtrsim \frac{\sqrt{\epsilon}}{r^2 \omega^2} \ll 1 \,,
\end{align}
since the numerator has a factor of $f^{1/2}\sim \epsilon^{1/2}$ and the remaining pieces tend to cause the denominator to be even smaller.
This approximation breaks down near the turning point $\omega^2 = V$, and there a matching calculation is used to get the final result~\eqref{eq:WKBWaveform}. 

The idea is to match this interior wavefunction to the exterior at the surface.
Provided that $\epsilon = 1 - \xi \ll1$ so that the stellar surface is close to $r = 2M$, we know that the wave solutions outside the ECO are freely propagating left and right waves in terms of the tortoise coordinate,
\begin{align}
    u_{\ell m \omega} = B[e^{-i \omega(r_* - R_*)} + \R e^{i \omega (r_* - R_*)}] \,. 
\end{align}
We have chosen the parametrization here to isolate the reflection coefficient $\R$, which multiplies the rightward traveling wave, and defined $R_*$ as the tortoise coordinate of the surface of the ECO.

Our goal is to compute $\R$ given the interior metric of the ECO.
For this, we match the wavefunction and its derivative at the surface.
For the exterior solution, we have
\begin{align}
    \left. \frac{d u/dr_*}{u} \right|_R = i \omega \frac{\R -1}{\R+1}\,.
\end{align}
For the interior solution, we have
\begin{align}
    \frac{du}{dr_*} & 
    = A \cos(\varphi - \pi/4) \frac{d}{dr_*} \left(\sqrt{\omega^2 - V}\right)^{-1/2}  
    +  \frac{i A}{2(\omega^2 - V)^{1/4}} \frac{d \varphi}{dr_*}\left(e^{i \varphi - i \pi/4} - e^{-i\varphi + i \pi/4}\right)
\end{align}
Now, $\sqrt{\omega^2 - V} = d\varphi/dr_*$, so the first term is
\begin{align}
    -\frac{A}{2} \left(\sqrt{\omega^2 - V}\right)^{-3/2} \frac{d^2\varphi}{dr_*^2} \cos(\varphi - \pi/4) 
    =
    -\frac{1}{2} \frac{d^2 \varphi/dr_*^2}{(d\varphi/dr_*)} u = - \frac{1}{2} \beta \sqrt{\omega^2 - V} \, u \,, 
\end{align}
and we neglect it in the WKB limit.
This gives for the inner solution
\begin{align}
    \left. \frac{d u/dr_*}{u} \right|_R \approx 
    i \left.\frac{d \varphi}{dr_*}\right|_R
    \frac{e^{i \varphi_R - i \pi/4} - e^{-i\varphi_R + i \pi/4}}{e^{i \varphi_R - i \pi/4} + e^{-i\varphi_R + i \pi/4}} \,,
\end{align}
Now, $d\varphi/dr_*|_R = [\omega^2 - V(R)]^{1/2} = \omega + O(\epsilon)$ under our assumptions, so the matching reduces to
\begin{align}
    i \omega \frac{\R -1}{\R+1} \approx i\omega \frac{e^{2i \varphi_R - i \pi/2} - 1}{e^{2i \varphi_R - i \pi/2} + 1} \,.
\end{align}
From this $\R$ can be read off and is given by Eq.~\eqref{eq:ECORef}.

\begin{figure}[t]
\includegraphics[width=0.49\columnwidth]{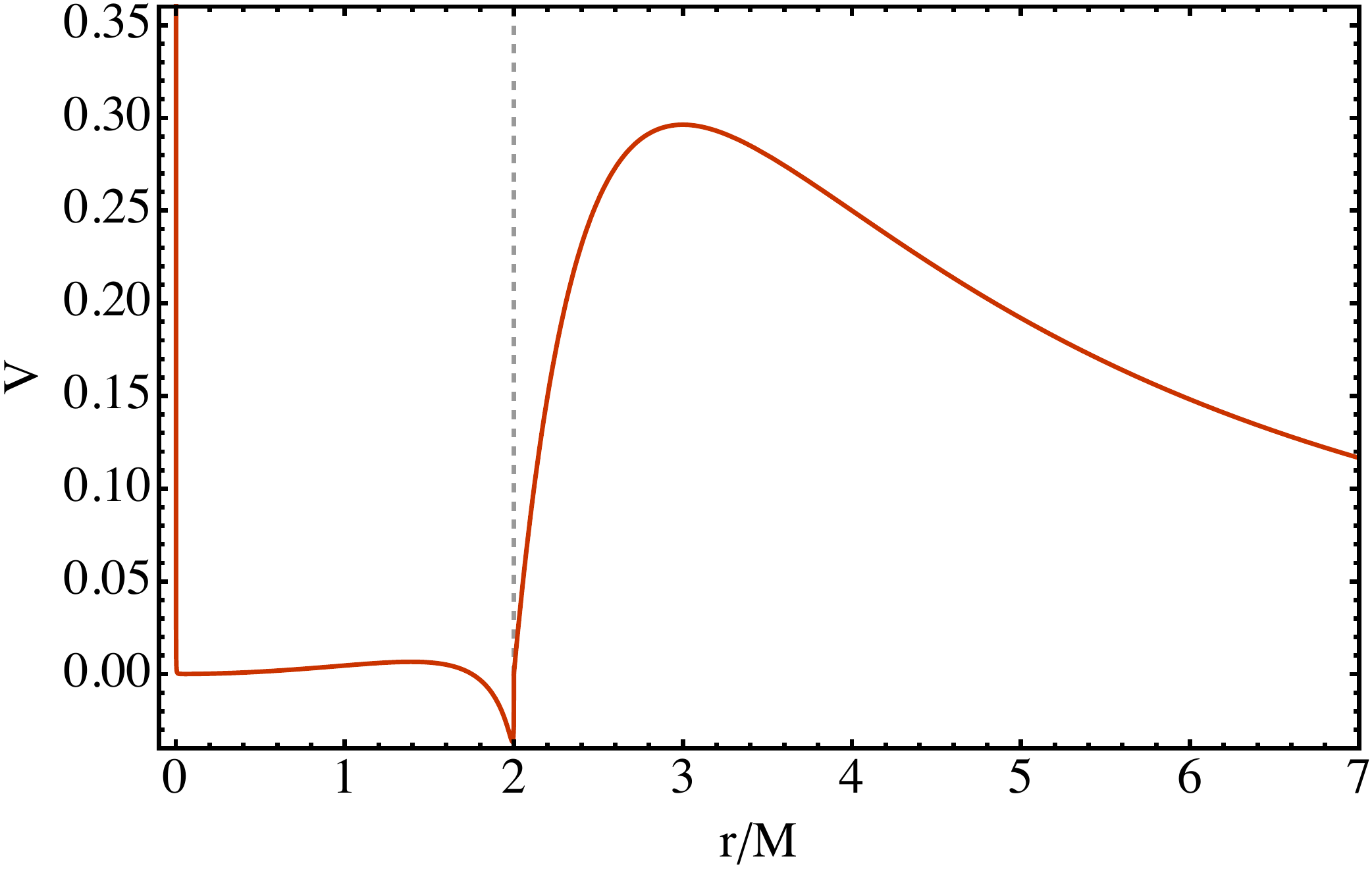}\\
\caption{Effective potential $V(r)$ for scalar wave propagation in the Buchdahl solution, with $\ell = 2$ and $\epsilon = 10^{-4}$.
}
\label{fig:BuchPotential}
\end{figure}

As an example, Fig.~\ref{fig:BuchPotential} shows this situation for Buchdahl's star, in the case $\epsilon = 10^{-4}$, where the surface of the star is at $2M(1+\epsilon)$.
The potential is relatively flat and small throughout the interior, until the sharp centrifugal barrier near the origin.
A small peak in the interior potential means that there is a minimum frequency below which the WKB analysis is invalid.
Just inside of the surface is a small region with $V < 0$, which presents no issues for the wave propagation for waves with positive frequencies, but may be an indication of the unphysical nature of this solution.
Negative energy solutions may exist in this region, but such waves would not be able to tunnel out of the star, since $V(r)>0$ for $r>R$.

\twocolumngrid

\bibliography{main.bbl}

\end{document}